%% file: delta_gamma.tex
\documentclass[preprint,12pt,superscriptaddress]{revtex4}  

\usepackage{graphics}
\usepackage[usenames]{color}
\usepackage{epsfig}

\def\cp{$CP$\/}

\def\mevm{~MeV/$c^2$\/}
\def\mevp{~MeV/$c$\/}
\def\meve{~MeV}
\def\gevm{~GeV/$c^2$\/}

\def\geve{~GeV}

\def\ra{\!\rightarrow\!}

\def\bbar{\overline{B}{}^{\,0}}

\def\bsdsds{$B^0_s\ra D^{(*)+}_s D^{(*)-}_s$}
\def\bsdspi{$B^0_s\ra D^{(*)-}_s \pi^+$}
\def\bdsd{$B^0\ra D^{(*)+}_s D^-$}

\def\bs{B^{}_s}
\def\bsst{B^{*}_s}
\def\bsbar{\overline{B}{}^{}_s}
\def\bsbarst{\overline{B}{}^{\,*}_s}
\def\fl{f^{}_L}

\def\mbc{M^{}_{\rm bc}}
\def\de{\Delta E}
\def\qq{$q\bar{q}$}

\def\dgs{\Delta\Gamma^{}_s}
\def\dgcp{\Delta\Gamma^{CP}_s}
\def\gs{\Gamma^{}_s}
\def\kstz{\overline{K}{}^{\,*0}}
\def\kstp{K^{*+}}
\def\qqbar{$q\bar{q}$}

\def\simge{\mathrel{%
   \rlap{\raise 0.511ex \hbox{$>$}}{\lower 0.511ex \hbox{$\sim$}}}}
\def\simle{\mathrel{
   \rlap{\raise 0.511ex \hbox{$<$}}{\lower 0.511ex \hbox{$\sim$}}}}

\begin{document}

\title{\large
\begin{flushright}
{\normalsize KEK Preprint 2010-13} \\
\vspace*{-0.15in}
{\normalsize UCHEP-10-03} 
\end{flushright}
\boldmath{
Observation of \bsdsds\ using $e^+e^-$ collisions 
and a determination of the $\bs$-$\bsbar$ 
width difference $\dgs$  
}}

\input{author_list.tex}

\begin{abstract}
We have made the first observation of \bsdsds\ decays
using 23.6~fb$^{-1}$ of data recorded by the Belle experiment
running on the $\Upsilon(5S)$ resonance. The branching fractions
are measured to be
${\cal B}(B^0_s\ra D^+_s D^-_s)=(1.03\,^{+0.39}_{-0.32}\,^{+0.26}_{-0.25})\%$,
${\cal B}(B^0_s\ra D^{*\pm}_s D^{\mp}_s)=(2.75\,^{+0.83}_{-0.71}\,\pm 0.69)\%$, 
and
${\cal B}(B^0_s\ra D^{*+}_s D^{*-}_s)=(3.08\,^{+1.22}_{-1.04}\,^{+0.85}_{-0.86})\%$;
the sum is 
${\cal B}(B^0_s\ra D^{(*)+}_s D^{(*)-}_s)=(6.85\,^{+1.53}_{-1.30}\,^{+1.79}_{-1.80})\%$.
Assuming \bsdsds\ saturates decays to \cp-even final states,  
the branching fraction determines the ratio $\dgs/\cos\varphi$, 
where $\dgs$ is the difference in widths between the two 
$\bs$-$\bsbar$ mass eigenstates, and $\varphi$ is a 
\cp-violating weak phase. Taking \cp\ violation
to be negligibly small, we obtain
$\dgs/\gs = 
0.147\,^{+0.036}_{-0.030}\,{\rm (stat.)}\,^{+0.042}_{-0.041}\,{\rm (syst.)}$,
where $\gs$ is the mean decay width. 
\end{abstract}


\maketitle

Decays of $\bs$ mesons help elucidate the weak 
Cabibbo-Kobayashi-Maskawa structure of the 
Standard Model (SM). Because they are not produced in 
$\Upsilon(4S)$ decays, $\bs$ mesons are much less 
studied than their $B^0_d$ and $B^\pm$ counterparts. 
Most $\bs$ data 
comes from the hadron collider experiments CDF and D\O.
Recently, another method to study $\bs$ decays has been 
exploited: that of running an $e^+e^-$ collider at 
a center-of-mass (CM) energy corresponding to the 
$\Upsilon(5S)$ resonance, which subsequently decays 
to $B^{(*)}_s\overline{B}{}^{(*)}_s$ pairs. Both the
CLEO~\cite{bs_cleo} and Belle~\cite{drutskoy07,bs_belle,remi} 
Collaborations have used this method to measure inclusive 
and exclusive $B^0_s$ decays. 
In this paper we use $L^{}_{\rm int}=23.6$~fb$^{-1}$ 
of data recorded by Belle at the $\Upsilon(5S)$ 
($\sqrt{s}=10.87$\geve) to make the first observation 
($>\!5\sigma$) of \bsdsds~\cite{charge-conjugates}.
First measurements of the $D^{(*)+}_sD^{(*)-}_s$ final state
were made by the ALEPH~\cite{ds_aleph} and D\O~\cite{ds_dzero}
Collaborations using inclusive $\phi\phi X$ and $D^+_sD^-_sX$ 
samples. CDF~\cite{ds_cdf} measured the single decay 
$B^0_s\ra D^+_sD^-_s$. Here we exclusively reconstruct 
all three final states:
$D^+_sD^-_s,\,D^{*\pm}_sD^\mp_s$, and $D^{*+}_sD^{*-}_s$.

These final states are expected to be predominantly 
\cp-even~\cite{Aleksan}, and the (Cabibbo-favored)
partial widths should dominate the difference in 
decay widths $\dgcp$ between the two $\bs$-$\bsbar$ 
\cp\ eigenstates~\cite{Aleksan}. 
This parameter equals $\dgs/\cos\varphi$, 
where $\dgs$ is the width difference between the mass eigenstates, 
and $\varphi$ is a \cp-violating weak phase~\cite{Dunietz}. Thus 
the branching fraction gives a constraint in $\dgs$-$\varphi$ 
parameter space. Both of these parameters 
can receive contributions from new physics~\cite{Nierste,new_physics}.
The values favored by current measurements~\cite{d0cdf_dg_avg} 
differ somewhat from the SM prediction~\cite{Nierste}.

The Belle detector~\cite{belle_detector} running at the
KEKB collider~\cite{kekb} includes a silicon vertex detector,
a central drift chamber, an array of aerogel threshold Cherenkov 
counters, time-of-flight scintillation counters, and an 
electromagnetic calorimeter. At the $\Upsilon(5S)$ resonance, 
the $e^+e^-\ra b\bar{b}$ cross section is
$\sigma^{}_{b\bar{b}}=0.302\pm 0.014$~nb~\cite{drutskoy07,bs_cleo},
and the fraction of $\Upsilon(5S)$ decays producing $\bs$ mesons 
is $f^{}_s=0.193\,\pm 0.029$~\cite{pdg}. Three production modes 
are kinematically allowed: $\bs\bsbar$, $\bs\bsbarst$ or $\bsst\bsbar$, 
and $\bsst\bsbarst$. In this analysis we use only the last 
(dominant) mode, for which the fraction is
$f^{}_{B^*_s\overline{B}{}^{\,*}_s}=0.901\,^{+0.038}_{-0.040}$~\cite{remi}.
The $\bsst$ decays via $\bsst\ra\bs\gamma$, and the $\gamma$
is not reconstructed. Thus the number of $\bs\bsbar$ pairs 
used in this analysis is 
$N^{}_{\bs\bsbar} = 
L^{}_{\rm int} \cdot \sigma^{}_{b\bar{b}} \cdot f^{}_s \cdot
f^{}_{B^*_s\overline{B}{}^{\,*}_s} = (1.24\pm 0.20)\times 10^6$.

We select $B^0_s\ra D^{*+}_s D^{*-}_s$, $D^{*\pm}_s D^{\mp}_s$, 
and $D^{+}_s D^{-}_s$ decays in which 
$D^+_s\ra \phi\pi^+$,
$K^0_S\,K^+$,
$\kstz K^+$,
$\phi\rho^+$,
$K^0_S\,\kstp$, and 
$\kstz \kstp$.
We require that charged tracks 
originate from near the $e^+e^-$ interaction point. 
Charged kaons are selected by requiring that a kaon likelihood
variable based on $dE/dx$ measured in the central drift chamber 
and information from the aerogel threshold Cherenkov counters
and time-of-flight scintillation counters be $>\!0.60$; this 
requirement is $\sim\!90$\% efficient and has a $\pi^\pm$ 
misidentification rate of $\sim\!10$\%.
Tracks having kaon likelihood $<\!0.60$ are identified 
as $\pi^\pm$. Neutral $K^0_S$ candidates are reconstructed 
from $\pi^+\pi^-$ pairs having an invariant mass within 
10\mevm\ of the $K^0_S$ mass~\cite{pdg} and satisfying 
loose requirements on the decay vertex position~\cite{goodKS}. 
The momentum of tracks (except the $\pi^\pm$ 
from $K^0_S$ decay) must be $>\!100$\mevp. 

Neutral $\pi^0$ candidates are reconstructed from $\gamma\gamma$ 
pairs having an invariant mass within 15\mevm\ of the
$\pi^0$ mass. The photons must have a laboratory energy 
greater than 100\meve.
Neutral $\kstz$ (charged $\kstp$)
candidates are reconstructed from a $K^-$ ($K^0_S$)
and $\pi^+$ having an invariant mass within 50\mevm\ of 
$M^{}_{K^{*0}}\,(M^{}_{K^{*+}})$.
Neutral $\phi$ (charged $\rho^+$) candidates are 
reconstructed from a $K^+K^-$ ($\pi^+\pi^0$) pair
having an invariant mass within 12\mevm\ (100\mevm) 
of $M^{}_{\phi}\,(M^{}_{\rho^+})$.

The invariant mass windows used for $D^+_s$ candidates 
are 10\mevm\ ($2.5\!-\!3.2\sigma$) for the three 
final states containing $K^*$ candidates, 
20\mevm\ ($1.7\sigma$) for $\phi\rho^+$, and
15\mevm\ ($\simge 4.0\sigma$) for the remaining 
two modes. For the three vector-pseudoscalar final 
states, we impose a loose
requirement on the helicity angle
$\theta^{}_{\rm hel}$, which is the angle between
the momentum of the charged daughter of the vector particle 
and the direction opposite the $D_s$ momentum in 
the rest frame of the vector particle. We require 
$|\cos\theta^{}_{\rm hel}|>0.20$, which retains 
99\% of signal decays and rejects 18\% of 
remaining background.

To reconstruct $D^{*+}_s\ra D^+_s\gamma$ decays,
we pair $D^+_s$ candidates with photon candidates
and require that the mass difference
$M^{}_{\tilde{D}_s^+\gamma} - M^{}_{\tilde{D}_s^+}$
be within 12.0\mevm\ of the nominal value (143.8\mevm), 
where $\tilde{D}_s^+$ denotes the reconstructed
$D^+_s$ candidate. 
This requirement (and that for ${\cal R}$ discussed below)
is determined by optimizing a figure-of-merit $S/\sqrt{S+B}$,
where $S$ is the expected signal based on Monte Carlo (MC) 
simulation and $B$ is the expected background as estimated 
from a data sideband.
We require that the photon energy in the CM system 
be greater than 50\meve, and that the energy deposited in 
the central $3\times 3$ array of cells of the 
electromagnetic calorimeter cluster be at least 
85\% of the energy deposited in the central 
$5\times 5$ array of cells.

Signal $\bs$ decays are reconstructed from $D^{(*)}_s D^{(*)}_s$ 
pairs using two quantities: the beam-energy-constrained mass 
$\mbc=\sqrt{E^2_{\rm beam} - p^2_B}$, and the energy difference
$\de= E^{}_B-E^{}_{\rm beam}$, where $p^{}_B$ and $E^{}_B$ are 
the reconstructed momentum and energy of the $B^0_s$, 
and $E_{\rm beam}$ is the beam energy. These quantities 
are evaluated in the $e^+e^-$ CM frame. 
When the $B^0_s$ is not fully reconstructed, e.g., due 
to losing the $\gamma$ from $D^{*+}_s\ra D^+_s\gamma$, 
$\de$ is shifted lower but $\mbc$ remains almost 
unchanged. We determine our signal yields by 
fitting events in the region
$5.20\mbox{\gevm}<\mbc <5.45$\gevm\ 
and $-0.15\mbox{\geve}<\de <0.10$\geve.
The modes
$\Upsilon(5S)\ra\bs\bsbar$, $\bs\bsbarst$ and $\bsst\bsbarst$
are well-separated in $\mbc$-$\de$ space.
We see no evidence for $\bs\bsbar$ and $\bs\bsbarst$ and
thus do not fit for them. The expected yields based on 
Ref.~\cite{remi} are less than one event for each of
$D_s^+D_s^-$, $D_s^{*\pm}D_s^{\mp}$, $D_s^{*+}D_s^{*-}$
final states.

Approximately half of the events have multiple \bsdsds\ candidates,
which usually arise from low momentum $\gamma$'s produced from $\pi^0$ 
decays. For these events we select the candidate that minimizes 
the quantity
\begin{eqnarray}
\chi^2 & = & \frac{1}{(2+N)}\,\biggl\{
\sum_{\#D^{}_s} \left[(\tilde{M}^{}_{D_s} - M^{}_{D_s})/\sigma^{}_M\right]^2 
\ +\  
\sum_{\#D_s^{*}} \left[(\widetilde{\Delta M} - 
\Delta M)/\sigma^{}_{\Delta M}\right]^2\biggr\}\,,
\end{eqnarray}
where $\Delta M=M^{}_{D^*_s}-M^{}_{D^{}_s}$;
$\tilde{M}^{}_{D^{}_s}$ and 
$\widetilde{\Delta M}$ 
are reconstructed quantities;
$\sigma^{}_M$ and $\sigma^{}_{\Delta M}$ are the 
uncertainties on $\tilde{M}^{}_{D^{}_s}$ and $\widetilde{\Delta M}$;
and the summations 
run over the two $D^+_s$ daughters and possible
$D^{*+}_s$ daughters ($N\!=\!0,1,2$) of a $B^0_s$ candidate. 
According to the MC simulation, this criterion selects 
the correct $B^0_s$ candidate 85\%, 76\%, and 75\% of the time, 
respectively, for $D^+_s D^-_s$, $D^{*\pm}_s D^{\mp}_s$, and 
$D^{*+}_s D^{*-}_s$ final states.

We reject background from 
$e^+e^-\ra q\bar{q}~(q=u,d,s,c)$ continuum events
based on event topology: \qq\ events tend to be collimated,
while $B^{}_{(s)}\overline{B}{}^{}_{(s)}$ events tend to be 
spherical. We distinguish these topologies 
using a  Fisher discriminant based on a set of 
modified Fox-Wolfram  moments~\cite{KSFW}. 
This discriminant is used to calculate
a likelihood $\mathcal{L}_{s}$ ($\mathcal{L}_{q\overline q}$)
for an event assuming the event is signal ($q\overline{q}$ background).
We form the ratio 
$\mathcal{R}=\mathcal{L}_{s}/(\mathcal{L}_{s}+\mathcal{L}_{q\overline q})$
and require ${\cal R}\!>\!0.20$. This selection is 95\% efficient 
for signal decays and removes $>80$\% of \qqbar\ background.

\begin{figure}
\begin{center}
\vbox{
\epsfig{file=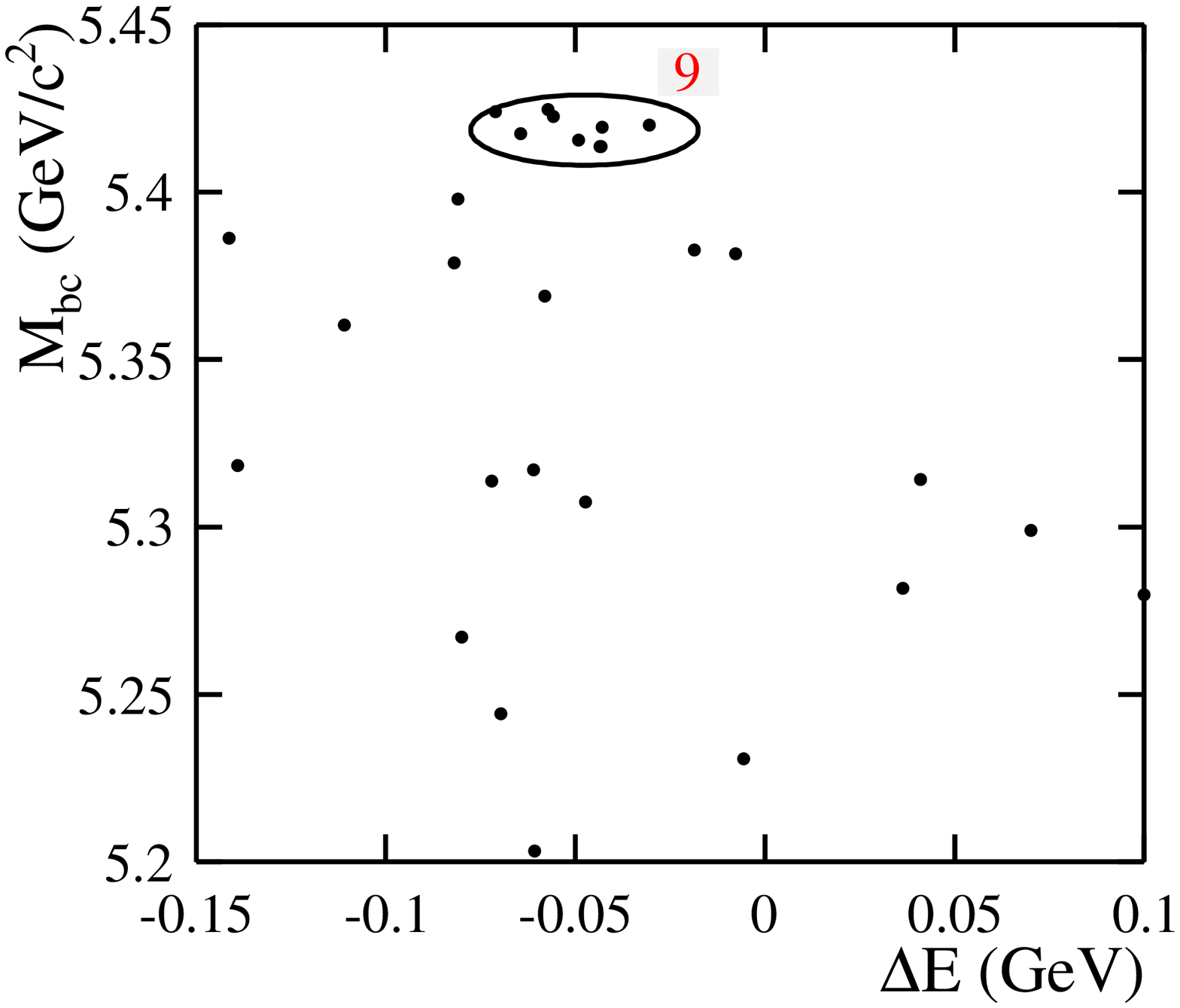,width=2.7in}
\vskip-0.26in
\epsfig{file=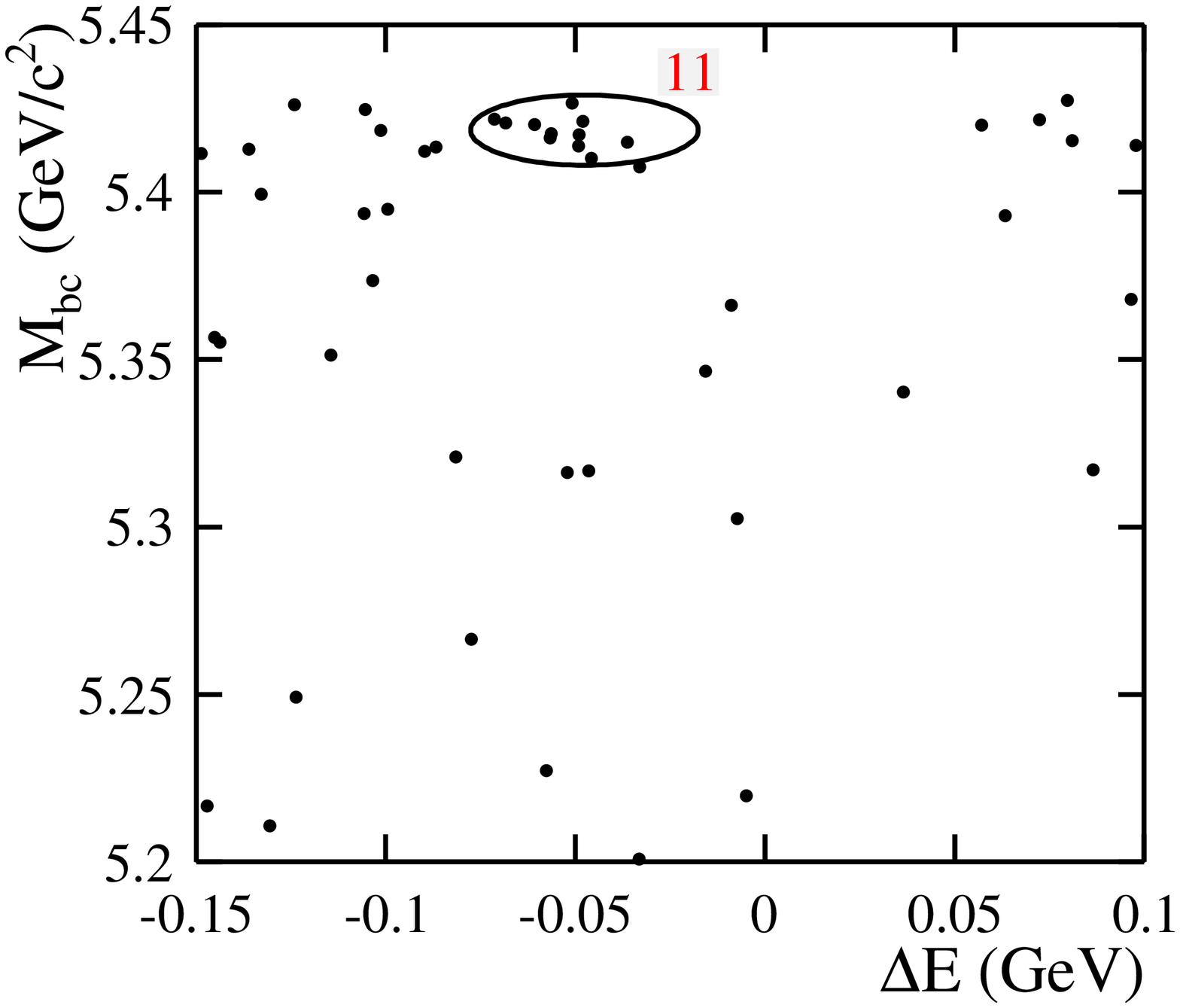,width=2.7in}
\vskip-0.26in
\epsfig{file=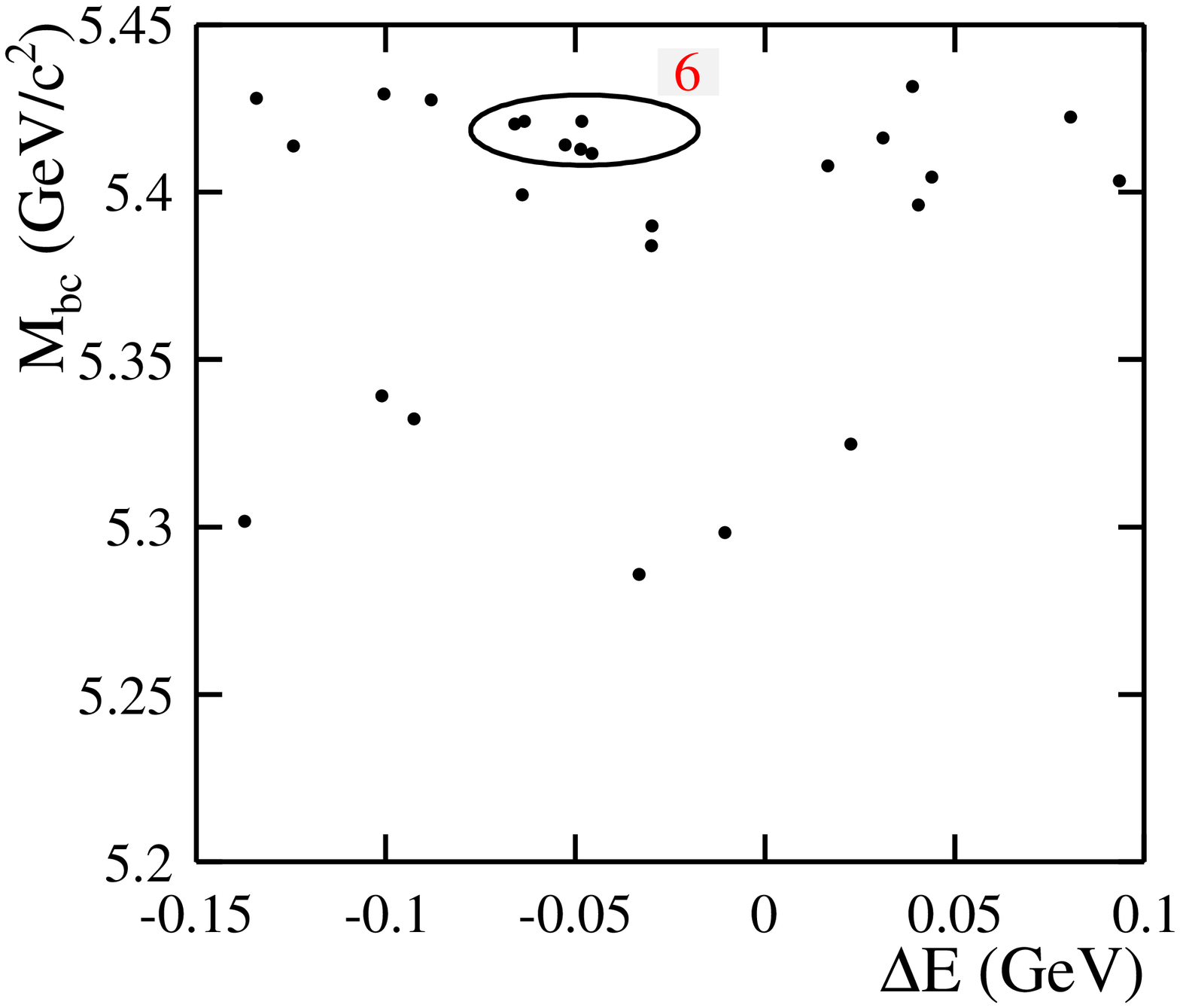,width=2.7in}
}
\end{center}
\vskip-0.20in
\caption{$\mbc$ vs. $\de$ scatter plots. The signal 
ellipses correspond to $3\sigma$ in resolution for
$\Upsilon(5S)\ra B^{*}_s \overline{B}{}^{*}_s$ decays;
the number of candidates within the ellipses is listed.
The top, middle, and bottom plots correspond to 
$B^0_s\ra D^+_s D^-_s$, $B^0_s\ra D^{*\pm}_s D^{\mp}_s$,
and $B^0_s\ra D^{*+}_s D^{*-}_s$, respectively. }
\label{fig:scatter_data}
\end{figure}

The remaining background consists of 
$\Upsilon(5S)\ra B^{(*)}_s\overline{B}{}^{(*)}_s\ra D^+_s X$,
$\Upsilon(5S)\ra BBX$ (where $b\bar{b}$ hadronizes into 
$B^0,\,\bbar$, or $B^\pm$), and
$B^{}_s\ra D^\pm_{sJ}(2317)D^{(*)}_s$,
$B^{}_s\ra D^\pm_{sJ}(2460)D^{(*)}_s$, and
$B^{}_s\ra D^\pm_{s}D^\mp_s\pi^0$ decays. The last three processes
peak at negative values of $\de$, and their yields are estimated
using analogous $B^{}_d\ra D^\pm_{sJ}D^{(*)}$ branching fractions.
The total yields for all backgrounds within an $\mbc$-$\de$ 
signal region spanning $3\sigma$ in $(\mbc,\de)$ resolution 
are $0.25\pm 0.03$, $0.25\pm 0.06$, and $0.15\pm 0.13$ events,
respectively, for $B^{}_s\ra D^+_s D^-_s$, $D^{*\pm}_s D^{\mp}_s$, 
and $D^{*+}_s D^{*-}_s$ decays.
To check our background estimates, we count events in 
the sideband region $\mbc<5.375$\gevm\ and find 
reasonable agreement with the yields predicted from 
MC simulation. All selection criteria are finalized 
before looking at events in the signal regions. 
The final event samples are
shown in Fig.~\ref{fig:scatter_data}.

To measure the signal yields, we perform a
two-dimensional extended unbinned maximum-likelihood
fit to the $\mbc$-$\de$ distributions. For each sample, 
we include probability density functions (PDFs) for 
signal and 
$q\bar{q}$, $B^{(*)}_s\overline{B}{}^{(*)}_s\ra D^+_s X$, 
and $\Upsilon(5S)\ra BBX$ backgrounds. As these
backgrounds have similar $\mbc$, $\de$ shapes, 
we use a single PDF for them, taken to be an ARGUS 
function~\cite{ARGUS} for $\mbc$ and a third-order 
Chebyshev polynomial for $\de$. All shape parameters 
are taken from MC simulation. Other backgrounds are
very small and considered only when evaluating
systematic uncertainties.

The signal PDFs have three components: correctly 
reconstructed (CR) decays; ``wrong combination'' (WC) 
decays in which a non-signal track or photon is included 
in place of a true daughter track or photon; and ``cross-feed'' (CF) 
decays in which a $D^{*\pm}_s D^{\mp}_s$ or $D^{*+}_s D^{*-}_s$ 
is reconstructed as
$D^+_s D^-_s$ or $D^{*\pm}_s D^{\mp}_s$, respectively, or
else a $D^+_s D^-_s$ or $D^{*\pm}_s D^{\mp}_s$ 
is reconstructed as $D^{*\pm}_s D^{\mp}_s$ or $D^{*+}_s D^{*-}_s$. 
In the former (latter) case the signal decay has lost
(gained) a photon, and $\de$ is typically shifted 
lower (higher) by 100-150\meve.
The PDF for CR events is modeled with a single Gaussian 
for $\mbc$ and a double-Gaussian with common mean for $\de$.
The means and widths are taken from MC simulation and 
calibrated using \bsdspi\ and \bdsd\ control samples. 
The PDFs for WC and CF events are modeled from 
MC simulation using non-parametric PDFs with Kernel 
Estimation~\cite{RooKeys}. 
The fractions of WC and CF-down events are also taken 
from MC simulation. The fractions of CF-up events 
are difficult to simulate and thus floated in
the fit, as the extraneous $\gamma$ usually originates 
from a $B$ decay chain and many $B, B^{}_s$ partial widths 
are unmeasured.
As the CF-down fractions
are fixed, the three distributions 
($D^+_s D^-_s,\ D^{*\pm}_s D^{\mp}_s$, and $D^{*+}_s D^{*-}_s$)
are fitted simultaneously~\cite{fit_errors}.
The CF fractions are typically 0.1--0.4.

\begin{figure}
\hskip0.2in
\vbox{
\hbox{
\epsfig{file=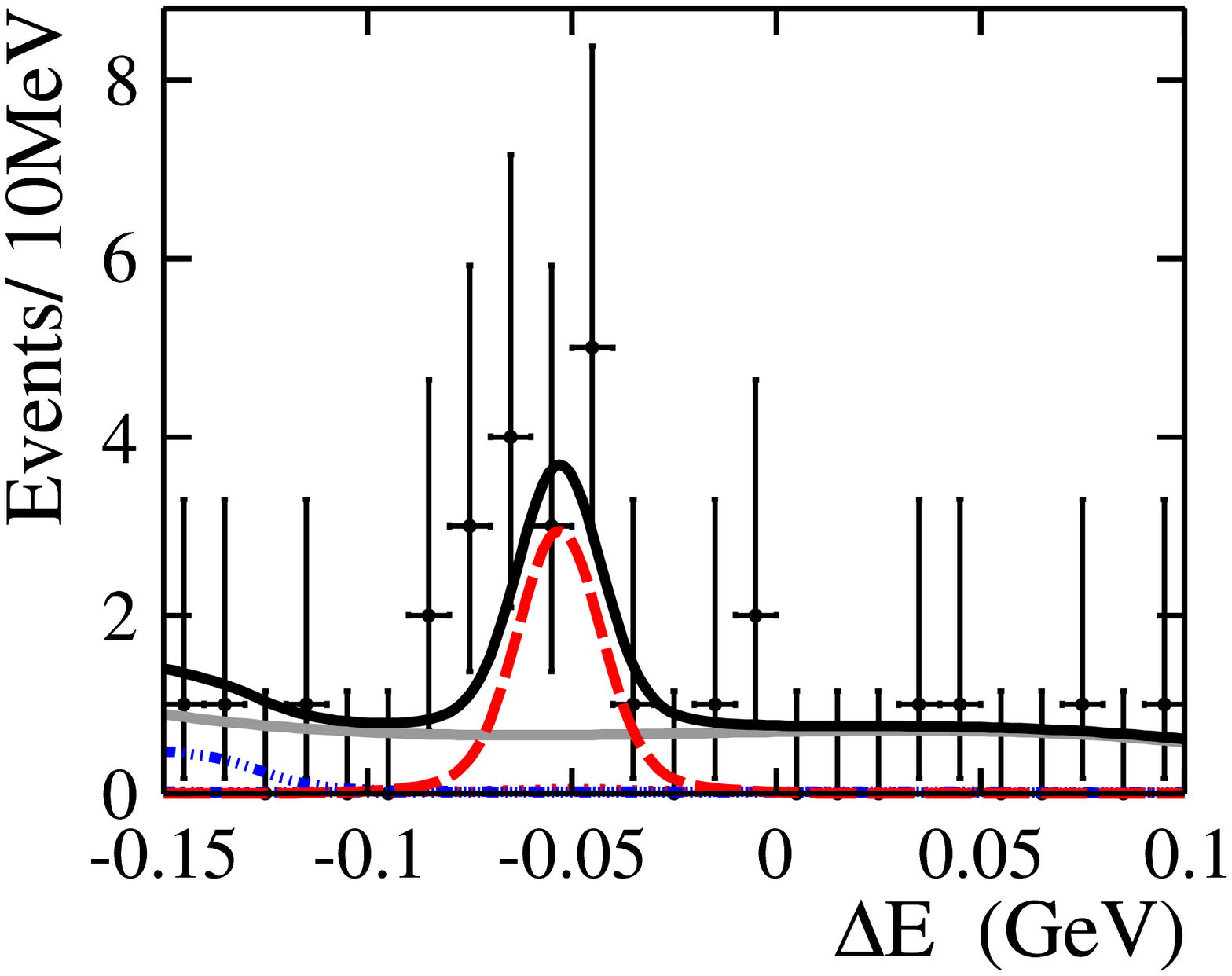,width=2.4in}
\hskip0.20in
\epsfig{file=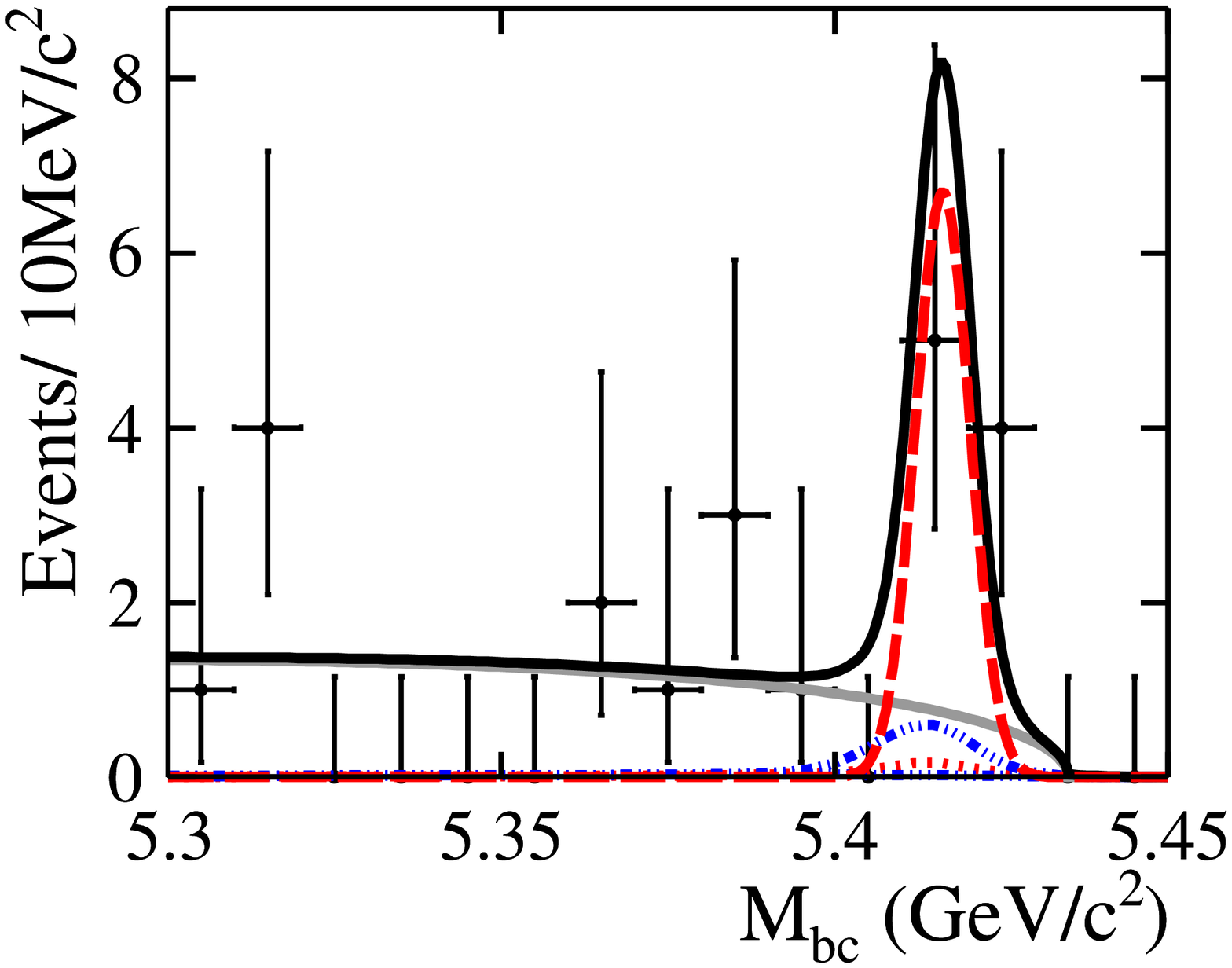,width=2.4in}
}
\hbox{
\epsfig{file=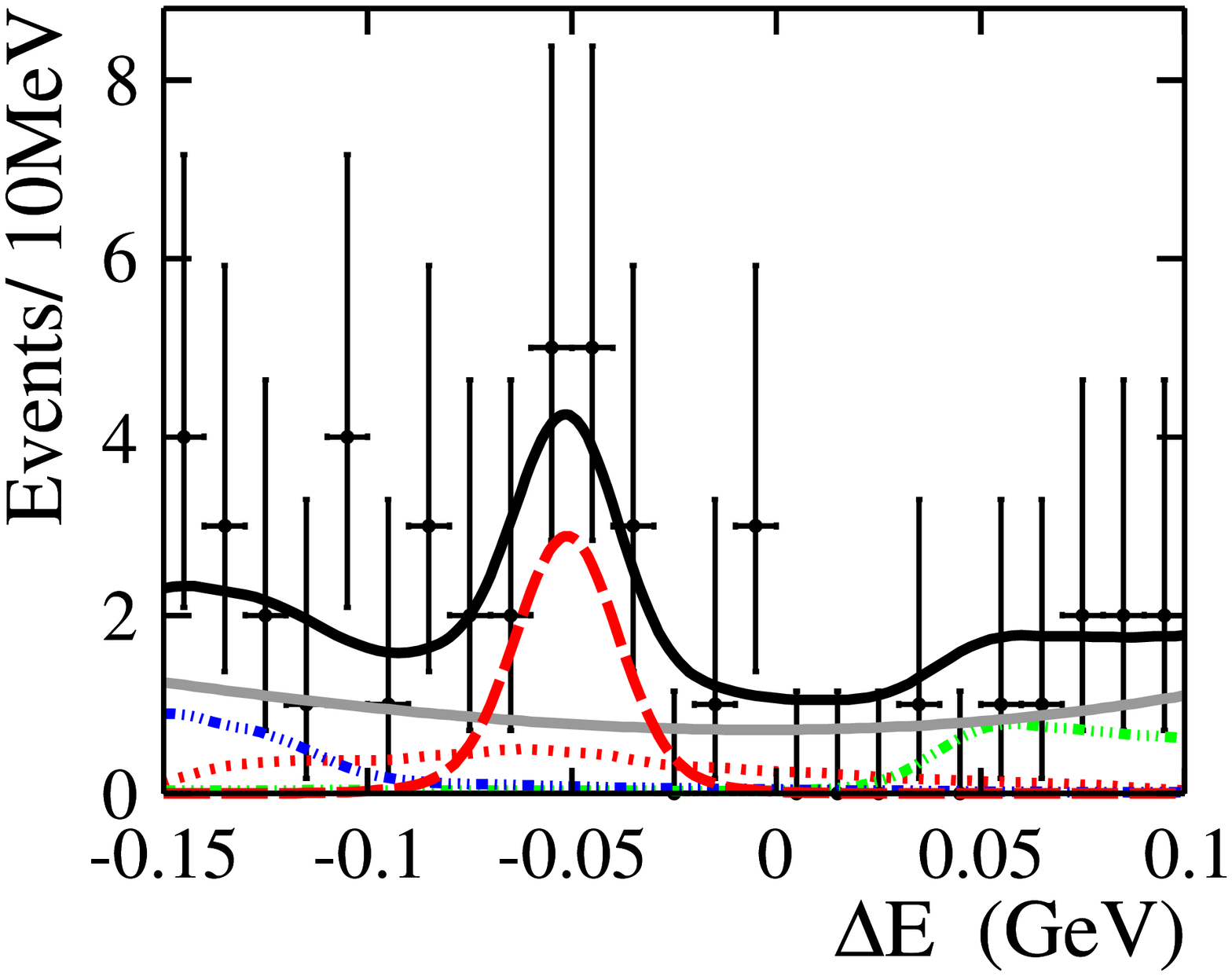,width=2.4in}
\hskip0.20in
\epsfig{file=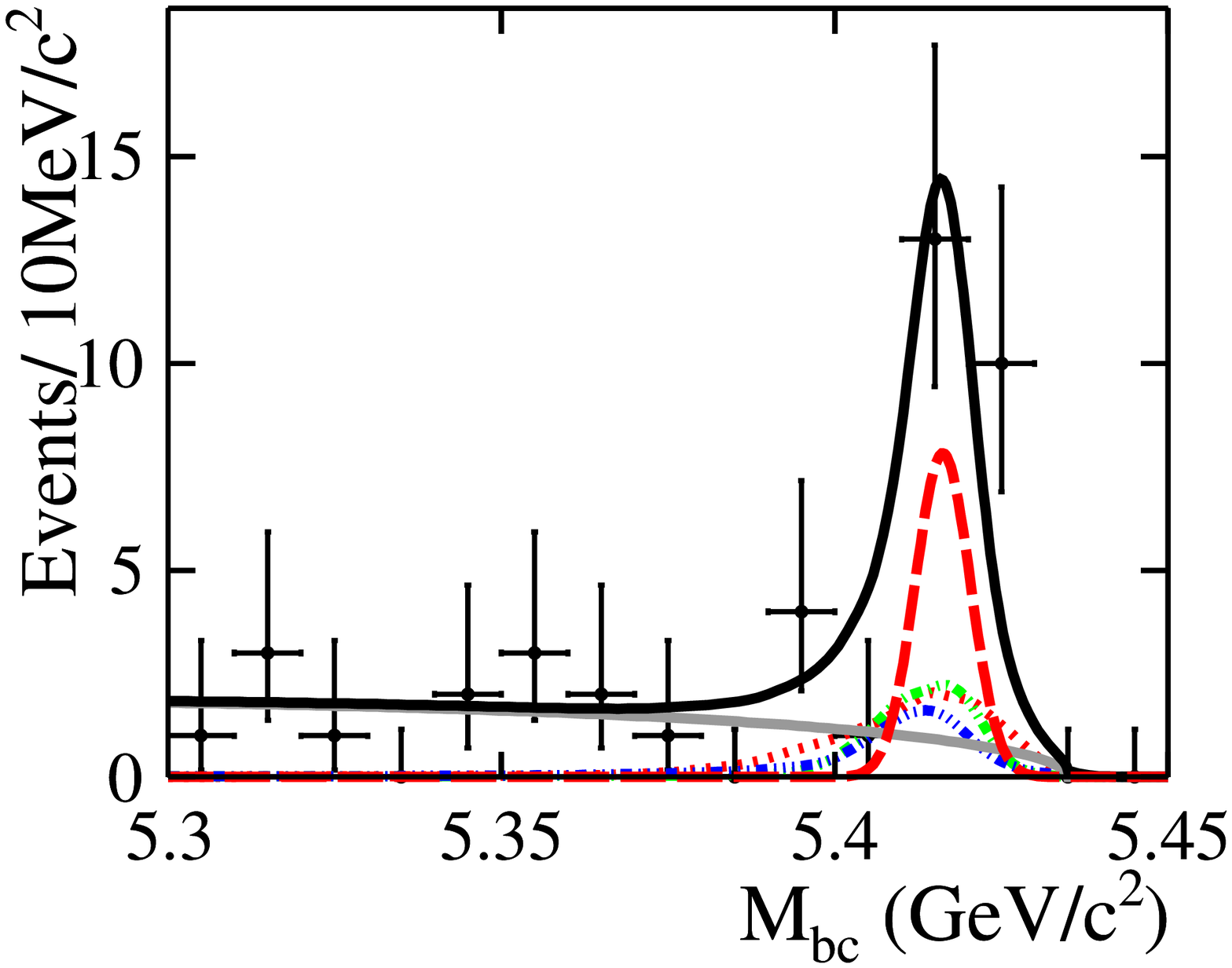,width=2.4in}
}
\hbox{
\epsfig{file=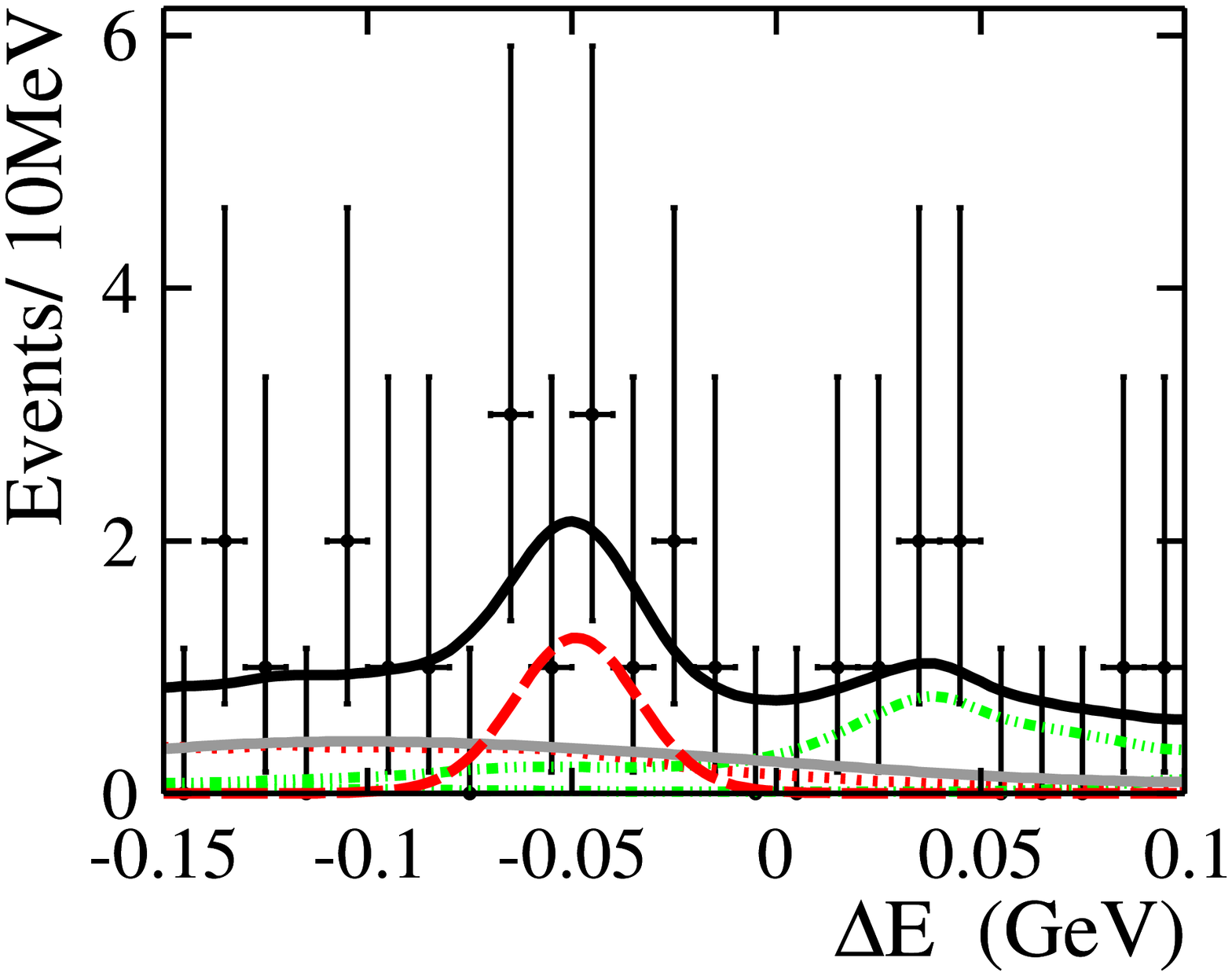,width=2.4in}
\hskip0.20in
\epsfig{file=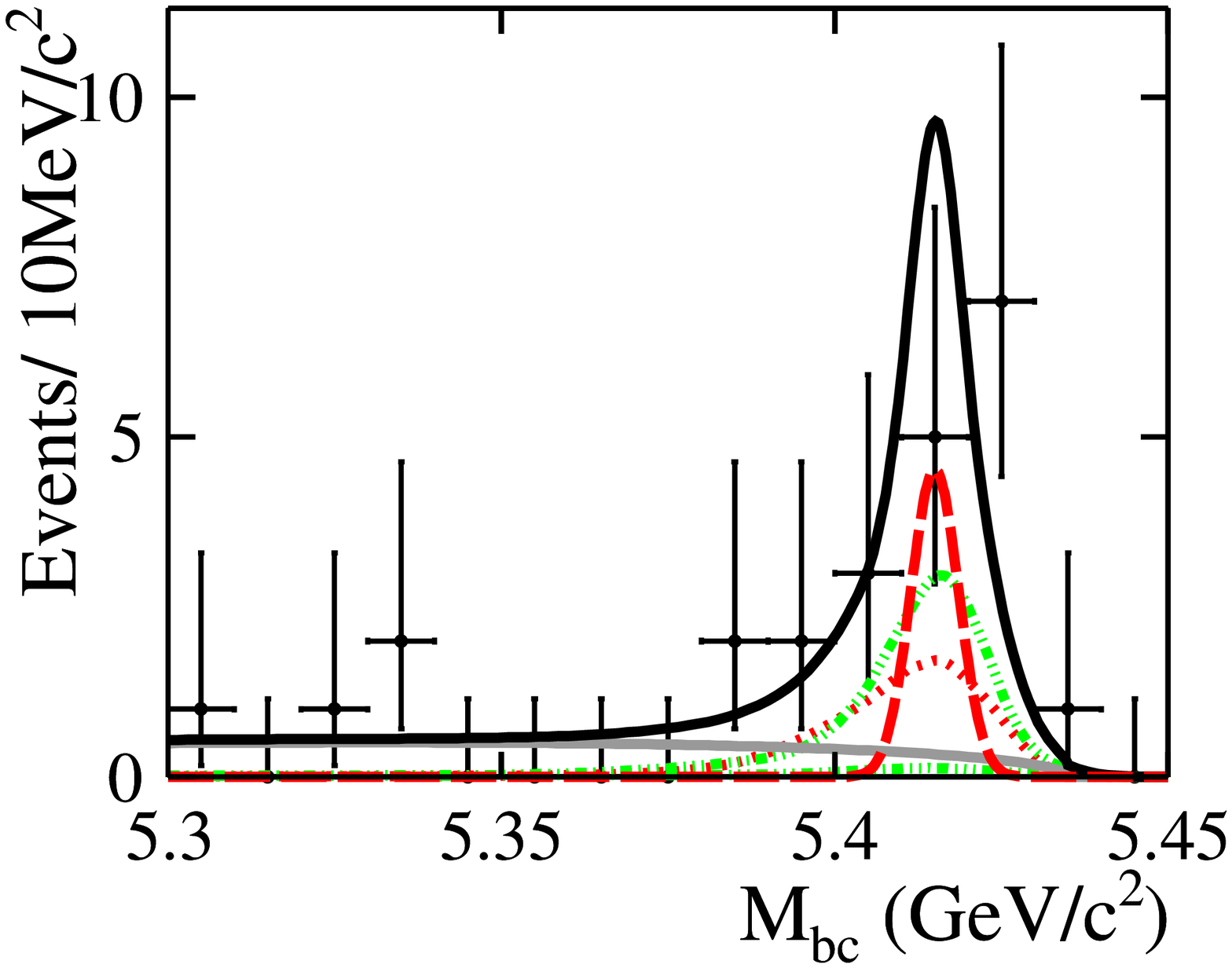,width=2.4in}
}
}
\caption{$\mbc$ and $\de$ projections of the fit result.
The rows correspond to $B^0_s\ra D^+_s D^-_s$ (top), 
$B^0_s\ra D^{*\pm}_s D^{\mp}_s$ (middle),and 
$B^0_s\ra D^{*+}_s D^{*-}_s$ (bottom).
The red dashed (dotted) curves show RC (WC) signal, 
the green and blue dash-dotted curves show CF signal, 
the grey solid curve shows background, and
the black solid curves show the total.
}
\label{fig:fit_results}
\end{figure}

The fit results are listed in Table~\ref{tab:fit_results},
and projections of the fit are shown in Fig.~\ref{fig:fit_results}.
The branching fraction for channel $i$ is calculated as
${\cal B}^{}_i = Y^{}_i/(\varepsilon^i_{MC}\cdot N^{}_{\bs\bsbar}\cdot 2)$, 
where $Y^{}_i$ is the fitted 
CR yield, and $\varepsilon^i_{MC}$ is the MC efficiency 
with intermediate branching fractions~\cite{pdg} included.
The efficiencies $\varepsilon^i_{MC}$ include small 
correction factors to account for differences between 
MC simulation and data for kaon identification. Inserting 
all values gives the branching fractions
listed in Table~\ref{tab:fit_results}.
The statistical significance is calculated
as $\sqrt{-2\ln(\mathcal{L}_0 / \mathcal{L}_{\mathrm{max}})}$, 
where $\mathcal{L}_0$ and $\mathcal{L}_{\mathrm{max}}$ are the  
values of the likelihood function when the signal yield $Y^{}_i$
is fixed to zero and when it is the fitted value, respectively. 
We include systematic uncertainty in the significance by smearing 
the likelihood function by a Gaussian having a width equal to 
the total systematic error related to the signal yield.

\begin{table}[htb]
\caption{Signal yields ($Y$), efficiencies 
including intermediate branching fractions ($\varepsilon$), 
branching fractions (${\cal B}$), and signal significance ($S$)
including systematic uncertainty. The first error listed is
statistical, the second is from systematics due to the
analysis procedure, and the third is from systematics 
due to external inputs. }
\begin{center}
\renewcommand{\arraystretch}{1.1}
\begin{tabular*}{0.80\textwidth}
{@{\extracolsep{\fill}} l | c c c c }
\hline\hline
Mode & $Y$ & $\varepsilon$ & ${\cal B}$ & $S$ \\
 & (events) & ($\times 10^{-4}$) & (\%) &   \\
\hline
$D^+_s D^-_s$ & 
$8.5\,^{+3.2}_{-2.6}$ & 3.31 & 
$1.03\,^{+0.39}_{-0.32}\,^{+0.15}_{-0.13}\,\pm 0.21$ & 6.2 \\
$D^{*\pm}_s D^{\mp}_s$ & 
$9.2\,^{+2.8}_{-2.4}$ & 1.35 & 
$2.75\,^{+0.83}_{-0.71}\,\pm 0.40\,\pm 0.56$ & 6.6 \\
$D^{*+}_s D^{*-}_s$ & 
$4.9\,^{+1.9}_{-1.7}$ & $0.643$ & 
$3.08\,^{+1.22}_{-1.04}\,^{+0.57}_{-0.58}\,\pm 0.63$ & 3.1 \\
\hline
Sum & 
$22.6\,^{+4.7}_{-3.9}$ &  & 
$6.85\,^{+1.53}_{-1.30}\,\pm 1.11\,^{+1.40}_{-1.41}$ & \\
\hline\hline
\end{tabular*} 
\end{center}
\label{tab:fit_results}
\end{table}

The systematic errors are listed in Table~\ref{tab:syst_errors}.
The error due
to PDF shapes is evaluated by varying shape parameters 
by $\pm 1\sigma$ for backgrounds, and by trying different 
parameterizations for the WC and CF components.
The systematic error for the fixed CF-down fractions
is evaluated by fitting a $B^0_s\ra D^{-}_s \pi^+$ 
control sample and comparing the fraction of 
$B^0_s\ra D^{*-}_s \pi^+$ contamination with that 
predicted by MC simulation. The difference 
is taken as the range of variation for the CF-down 
fractions.
We vary the fractions over this range and take
the r.m.s.~variation in ${\cal B}$ as the
systematic error. The error due to the fixed WC 
fractions is evaluated in the same way, but the
range of variation for the WC fractions is $\pm 20$\%.
The uncertainty due to $K^\pm$ identification depends 
on momentum but is $\sim$\,2.5\% per track; as our 
final states typically have four charged kaons, this 
error is 10-11\%.
The error due to tracking efficiency is 1.0\% per track. 
Uncertainty due to the unknown $B^0_s\ra D^{*+}_sD^{*-}_s$ 
longitudinal polarization fraction ($\fl$)
affects all three modes due to the CF components. 
For our nominal result, we take $\fl$ to be the 
world average (WA) value for the analogous spectator decay
$B^0_d\ra D^{*+}_s D^{*-}$: $0.52\pm\,0.05$~\cite{pdg}. 
The systematic error 
is taken as the change in ${\cal B}$ when $f^{}_L$ is 
varied by twice the error on the WA value.
Significant uncertainties
arise from $D^+_s$ branching fractions,
$\sigma^{}_{\Upsilon(5S)}$, $f^{}_s$, and 
$f^{}_{B^*_s\overline{B}{}^{\,*}_s}$, which are external 
factors that should be measured more precisely in the future. 
We list separately the systematic error due to these factors 
in Table~\ref{tab:fit_results}.

\begin{table}[htb]
\caption{Systematic errors (\%). The first twelve sources
affect the signal yield and thus the signal significance.}
\begin{center}
\renewcommand{\arraystretch}{0.9}
\begin{tabular*}{0.80\textwidth}
{@{\extracolsep{\fill}} l | c c | c c | c c }
\hline\hline
Source & 
\multicolumn{2}{c|}{$D^+_sD^-_s$} & 
\multicolumn{2}{c|}{$D^{*\pm}_s D^{\mp}_s$} & 
\multicolumn{2}{c}{$D^{*+}_s D^{*-}_s$} \\
 & $+\sigma$ & $-\sigma$ & $+\sigma$ & $-\sigma$ & $+\sigma$ & $-\sigma$ \\
\hline
CR PDF Shape             &  0.8 & 0.8 &  0.3 & 0.3 & 0.5 & 0.4  \\
Background PDF           &  1.1 & 1.3 &  1.9 & 2.0 & 3.0 & 6.1  \\
WC+CF  PDF               &  0.3 & 0.3 &  1.5 & 1.5 & 4.4 & 4.4  \\
WC/CF Fractions          &  0.2 & 0.2 &  5.0 & 5.0 & 8.7  & 8.7  \\
${\cal R}$ Requirement ($q\bar{q}$ suppr.)
    &  1.8 & 1.8 &  1.8 & 1.8 & 1.8  & 1.8  \\
Best Candidate Selection &  6.9 & 0.0 &  2.2 & 0.0 & 2.2  & 0.0  \\
$K^\pm$ Identification   & 10.1 & 10.1 & 10.6 & 10.6 & 10.9 & 10.9 \\
$K^0_S$ Reconstruction   &  2.1 &  2.1 & 2.1 & 2.1 &  2.2 &  2.2 \\
$\pi^{0}$ Reconstruction &  1.1 &  1.1 & 1.1 & 1.1 &  1.0 &  1.0 \\
$\gamma$                 &   -  &  -   & 3.8 & 3.8 &  7.6 &  7.6 \\
Tracking                 &  6.2 &  6.2 & 6.2 & 6.2 &  6.2 &  6.2 \\
Polarization             &  0.2 &  0.0 & 0.8 & 0.5  & 0.7 &  0.3 \\
\hline
MC Statistics for $\varepsilon$& 1.1 & 1.1 & 0.9 & 0.8 &  1.0 &  1.0 \\
$D_{s}^{(*)}$ Branching Fractions & 12.4 & 12.4 & 12.4 & 12.4 & 12.5 & 12.5 \\
Luminosity               & \multicolumn{6}{c}{$\pm 1.3$} \\
$\sigma^{}_{\Upsilon(5S)}$ & \multicolumn{6}{c}{$\pm 4.6$} \\
$f^{}_s$                  & \multicolumn{6}{c}{$\pm 15$} \\
$f^{}_{B^*_s\overline{B}^*_s}$ & \multicolumn{6}{c}{$^{+4.2}_{-4.4}$} \\
\hline
Total  & 24.9 & 24.0 & 25.1 & 25.1 & 27.5 & 28.0  \\
\hline\hline
\end{tabular*} 
\end{center}
\label{tab:syst_errors}
\end{table}

In the limit
$m^{}_{c,b}\ra\infty$ while $(m^{}_b-2m^{}_c)\ra 0$,
the $b\ra c\bar{c}s$ process saturates the decay width~\cite{Shifman}.
If also the number of colors $N^{}_c\ra\infty$, then
$B^0_s\ra D^{*+}_s D^{*-}_s,\,D^{*\pm}_s D^{\mp}_s$ (along with
$D^+_s D^-_s$) are $CP$ even ($+$), and 
$\Gamma[B^0_{s}(CP+)\ra D^{(*)}_sD^{(*)}_s]$ 
saturates $\dgcp$~\cite{Aleksan}. This gives the relationship
$2{\cal B}(B^0_s\ra D^{(*)+}_sD^{(*)-}_s) = (\dgcp/2)[(1+\cos\varphi)/\Gamma^{}_L + 
(1-\cos\varphi)/\Gamma^{}_H]$, where $\Gamma^{}_{L,H}$
are the decay widths of the light and heavy mass eigenstates~\cite{Dunietz}.
Substituting $\Gamma^{}_{L,H} = \Gamma\pm\dgs/2$ and
$\dgcp = \dgs/\cos\varphi$~\cite{Dunietz} allows one to use 
the branching fraction ${\cal B}$ to constrain $\dgs$ and $\varphi$.
If \cp\ violation is negligible, then $\cos\varphi\!\simeq\!1$ 
and the above expression can be inverted to give
$\dgs/\gs = 2{\cal B}/(1-{\cal B})$.
Inserting ${\cal B}$ from Table~\ref{tab:fit_results} yields
\begin{eqnarray}
\frac{\dgs}{\gs} & = & 
0.147\,^{+0.036}_{-0.030}\,^{+0.042}_{-0.041}\,,
\label{eqn:dg_result}
\end{eqnarray}
where the first error is statistical and the second is systematic. 
This result is $1.3\sigma$ higher than that 
of Ref.~\cite{ds_dzero} but consistent with the
theory prediction~\cite{Nierste}.
There is theoretical uncertainty
arising from the \cp-odd component in 
$B^0\ra D^{*\pm}_s D^{\mp}_s, D^{*+}_sD^{*-}_s$
and contributions from other two-body final states; 
the effect upon $\dgs/\gs$ is estimated in Ref.~\cite{Aleksan}
to be $\pm 3$\%. This is much smaller than the
statistical/systematic errors on our measurement, but
there may be additional contributions coming from three-body 
final states, which are neglected in~\cite{Aleksan}.

In summary, we have measured the branching fractions for
\bsdsds\ using $e^+e^-$ data taken at the $\Upsilon(5S)$
resonance. Our results 
constitute the first observation of $B^0\ra D^{*\pm}_s D^{\mp}_s$ 
($6.6\sigma$ significance) and provide the first evidence 
for $B^0_s\ra D^{*+}_s D^{*-}_s$ ($3.1\sigma$ significance).
We use these measurements to determine 
the $B^0_s$-$\bsbar$ decay width difference~$\dgs$
with improved fractional precision.

We thank R.\,Aleksan and L.\,Oliver for useful discussions.
We thank the KEKB group for excellent operation of the
accelerator, the KEK cryogenics group for efficient solenoid
operations, and the KEK computer group and
the NII for valuable computing and SINET3 network support.  
We acknowledge support from MEXT, JSPS and Nagoya's TLPRC (Japan);
ARC and DIISR (Australia); NSFC (China); MSMT (Czechia);
DST (India); MEST, NRF, NSDC of KISTI (Korea); MNiSW (Poland); 
MES and RFAAE (Russia); ARRS (Slovenia); SNSF (Switzerland); 
NSC and MOE (Taiwan); and DOE (USA).

\end{document}

%% file: author_list.tex
\affiliation{Budker Institute of Nuclear Physics, Novosibirsk}
\affiliation{Faculty of Mathematics and Physics, Charles University, Prague}
\affiliation{University of Cincinnati, Cincinnati, Ohio 45221}
\affiliation{Department of Physics, Fu Jen Catholic University, Taipei}
\affiliation{Justus-Liebig-Universit\"at Gie\ss{}en, Gie\ss{}en}
\affiliation{The Graduate University for Advanced Studies, Hayama}
\affiliation{Hanyang University, Seoul}
\affiliation{University of Hawaii, Honolulu, Hawaii 96822}
\affiliation{High Energy Accelerator Research Organization (KEK), Tsukuba}
\affiliation{Institute of High Energy Physics, Chinese Academy of Sciences, Beijing}
\affiliation{Institute of High Energy Physics, Vienna}
\affiliation{Institute of High Energy Physics, Protvino}
\affiliation{Institute for Theoretical and Experimental Physics, Moscow}
\affiliation{J. Stefan Institute, Ljubljana}
\affiliation{Kanagawa University, Yokohama}
\affiliation{Institut f\"ur Experimentelle Kernphysik, Karlsruher Institut f\"ur Technologie, Karlsruhe}
\affiliation{Korea Institute of Science and Technology Information, Daejeon}
\affiliation{Korea University, Seoul}
\affiliation{Kyungpook National University, Taegu}
\affiliation{\'Ecole Polytechnique F\'ed\'erale de Lausanne (EPFL), Lausanne}
\affiliation{Faculty of Mathematics and Physics, University of Ljubljana, Ljubljana}
\affiliation{University of Maribor, Maribor}
\affiliation{Max-Planck-Institut f\"ur Physik, M\"unchen}
\affiliation{University of Melbourne, School of Physics, Victoria 3010}
\affiliation{Nagoya University, Nagoya}
\affiliation{Nara Women's University, Nara}
\affiliation{National Central University, Chung-li}
\affiliation{National United University, Miao Li}
\affiliation{Department of Physics, National Taiwan University, Taipei}
\affiliation{H. Niewodniczanski Institute of Nuclear Physics, Krakow}
\affiliation{Nippon Dental University, Niigata}
\affiliation{Niigata University, Niigata}
\affiliation{Novosibirsk State University, Novosibirsk}
\affiliation{Osaka City University, Osaka}
\affiliation{University of Science and Technology of China, Hefei}
\affiliation{Seoul National University, Seoul}
\affiliation{Sungkyunkwan University, Suwon}
\affiliation{School of Physics, University of Sydney, NSW 2006}
\affiliation{Tata Institute of Fundamental Research, Mumbai}
\affiliation{Excellence Cluster Universe, Technische Universit\"at M\"unchen, Garching}
\affiliation{Tohoku Gakuin University, Tagajo}
\affiliation{Department of Physics, University of Tokyo, Tokyo}
\affiliation{Tokyo Metropolitan University, Tokyo}
\affiliation{Tokyo University of Agriculture and Technology, Tokyo}
\affiliation{IPNAS, Virginia Polytechnic Institute and State University, Blacksburg, Virginia 24061}
\affiliation{Yonsei University, Seoul}
 \author{S.~Esen}\affiliation{University of Cincinnati, Cincinnati, Ohio 45221} 
  \author{A.~J.~Schwartz}\affiliation{University of Cincinnati, Cincinnati, Ohio 45221} 
  \author{I.~Adachi}\affiliation{High Energy Accelerator Research Organization (KEK), Tsukuba} 
  \author{H.~Aihara}\affiliation{Department of Physics, University of Tokyo, Tokyo} 
\author{K.~Arinstein}\affiliation{Budker Institute of Nuclear Physics, Novosibirsk}\affiliation{Novosibirsk State University, Novosibirsk} 
  \author{V.~Aulchenko}\affiliation{Budker Institute of Nuclear Physics, Novosibirsk}\affiliation{Novosibirsk State University, Novosibirsk} 
  \author{T.~Aushev}\affiliation{\'Ecole Polytechnique F\'ed\'erale de Lausanne (EPFL), Lausanne}\affiliation{Institute for Theoretical and Experimental Physics, Moscow} 
  \author{T.~Aziz}\affiliation{Tata Institute of Fundamental Research, Mumbai} 
  \author{A.~M.~Bakich}\affiliation{School of Physics, University of Sydney, NSW 2006} 
  \author{V.~Balagura}\affiliation{Institute for Theoretical and Experimental Physics, Moscow} 
% ...
\collaboration{The Belle Collaboration}

%% file: delta_gamma.bbl
\begin{thebibliography}{99}
\bibitem{charge-conjugates} 
Charge-conjugate modes are implicitly included.
\bibitem{fit_errors} Thus the fit errors for yields are less than 
the square root of the yields due to the CF information.
\end{thebibliography}
